\documentclass[bimj,fleqn]{w-art}
\usepackage{times}
\usepackage{w-thm}
\usepackage[authoryear]{natbib}
\theoremstyle{plain}
\theoremstyle{definition}
\usepackage[]{graphicx}
\chardef\bslash=`\\ % p. 424, TeXbook

\usepackage[colorlinks,bookmarksopen,bookmarksnumbered,citecolor=blue,urlcolor=black]{hyperref}
\usepackage{url}
\usepackage{threeparttable}
\usepackage{multirow}
\usepackage{booktabs}
\usepackage[dvipsnames]{xcolor}

\newcommand{\bfX}{{\bf X}}

\newcommand{\bfY}{{\bf Y}}
\newcommand{\bfmu}{\mbox{\boldmath $\mu$}}

\newcommand{\bfbeta}{\mbox{\boldmath $\beta$}}

\newcommand{\bfone}{{\bf 1}}
\newcommand{\bfzero}{{\bf 0}}
\newcommand{\bfI}{{\bf I}}
\newcommand{\bfV}{{\bf V}}
\newcommand{\bfR}{{\bf R}}
\newcommand{\bfJ}{{\bf J}}
\newcommand{\bfA}{{\bf A}}

\newcommand{\bfD}{{\bf D}}
\newcommand{\bfC}{{\bf C}}

\newcommand{\bfQ}{{\bf Q}}
\newcommand{\bfSigma}{\mbox{\boldmath $\Sigma$}}
\newcommand{\bfalpha}{\mbox{\boldmath $\alpha$}}

\newcommand{\ii}{{ijkl}}
\newcommand{\bftab}{\fontseries{b}\selectfont}

\begin{document}
%\DOIsuffix{bimj.DOIsuffix}
%\DOIsuffix{bimj.200100000}
%\Volume{52}
%\Issue{61}
%\Year{2010}
%\pagespan{1}{}
\keywords{Cluster randomized trials; Eigenvalues; Extended nested exchangeable correlation; Matrix-adjusted estimating equations (MAEE); Sample size.\\[1pc]
\noindent\hspace*{-4.2pc} Supporting Information for this article is available from the author or at \newline \url{https://github.com/XueqiWang/Four-Level_CRT}.}  %%% semicolon and fullpoint added here for keyword style

\title[Sample size and power analysis for four-level CRTs]{Power considerations for generalized estimating equations analyses of four-level \textcolor{black}{cluster randomized trials}}
%% Information for the first author.
\author[Wang {\it{et al.}}]{Xueqi Wang\footnote{Corresponding author: {\sf{e-mail: xueqi.wang@duke.edu}}, Phone: +1-217-898-5630}\inst{,1,2}} 
\address[\inst{1}]{Department of Biostatistics and Bioinformatics, Duke University School of Medicine, Durham, NC, 27707, USA}
\address[\inst{2}]{Duke Global Health Institute, Durham, NC, 27707, USA}
%%%%    Information for the second author
\author[]{Elizabeth L. Turner\inst{1,2}}
%%%%    Information for the third author
\author[]{John S. Preisser\inst{3}}
\address[\inst{3}]{Department of Biostatistics, University of North Carolina at Chapel Hill, Chapel Hill, NC, 27599, USA}
%%%%    Information for the fourth author
\author[]{Fan Li\inst{4,5}}
\address[\inst{4}]{Department of Biostatistics, Yale University School of Public Health, New Haven, CT, 06511, USA}
\address[\inst{5}]{Center for Methods in Implementation and Prevention Science, Yale University, New Haven, CT, 06511, USA}
%%%%    \dedicatory{This is a dedicatory.}
%\Receiveddate{zzz} \Reviseddate{zzz} \Accepteddate{zzz} 

\begin{abstract}
In this article, we develop methods for sample size and power calculations in four-level intervention studies when intervention assignment is carried out at any level, with a particular focus on cluster randomized trials (CRTs). CRTs involving four levels are becoming popular in health care research, where the effects are measured, for example, from evaluations (level 1) within participants (level 2) in divisions (level 3) that are nested in clusters (level 4). In such multi-level CRTs, we consider three types of intraclass correlations between different evaluations to account for such clustering: that of the same participant, that of different participants from the same division, and that of different participants from different divisions in the same cluster. Assuming arbitrary link and variance functions, with the proposed correlation structure as the true correlation structure, closed-form sample size formulas for randomization carried out at any level (including individually randomized trials within a four-level clustered structure) are derived based on the generalized estimating equations approach using the model-based variance and using the sandwich variance with an independence working correlation matrix. We demonstrate that empirical power corresponds well with that predicted by the proposed method for as few as 8 clusters, when data are analyzed using the matrix-adjusted estimating equations for the correlation parameters with a bias-corrected sandwich variance estimator, under both balanced and unbalanced designs.
\end{abstract}

\maketitle

\section{Introduction}\label{sec:intro}
Cluster randomized trials (CRTs) are commonly used to study the effectiveness of health care interventions within a pragmatic trials framework \citep{Weinfurt2017}. In CRTs, the unit of randomization is typically a group (or cluster) of individuals with outcome measurements taken on individuals themselves. Reasons to randomize clusters rather than individuals include administrative convenience and prevention of intervention contamination \citep{murray1998,turner2017a}. CRTs typically include two levels, for example, where patients are nested within clinics that are randomized to intervention conditions. With an increase in the number of pragmatic trials embedded within health care systems, recent CRTs have been designed with multiple hierarchical levels involving, for example, patients, providers, and health care systems \citep{heo2008statistical,teerenstra2010,liu2020}. Moreover, depending on the nature of the intervention and of the context, the unit of randomization may be at the highest level or one of the lower levels in the hierarchy. The inherent hierarchical structure of the health care delivery system demands rigorous multi-level methods that enable a precise evaluation of health care interventions. 

\begin{figure}[htbp]
\centering
\includegraphics[scale=0.4]{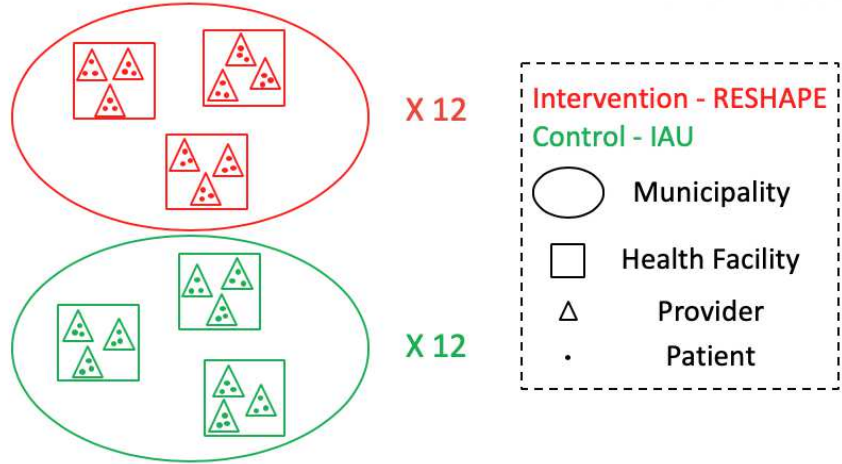}
\caption{An illustration of the RESHAPE trial with 12 municipalities per arm; the outcome of interest is whether patients are accurately diagnosed by provider (as determined by psychiatrist diagnosis). A motivating design question is: Given 3 health facilities per municipality, 3 providers per health facilities, and 3 patients per provider, how many municipalities are needed for a two-sided test to identify a difference in diagnostic accuracy of 88\% vs. 78\% at 90\% power given the 5\% significance level?}
\label{fig:Fig1}
\end{figure}

While methods for designing three-level CRTs have been previously developed in \citet{heo2008statistical}, \citet{teerenstra2010} and \citet{cunningham2016design}, there has been little development on methods for designing CRTs with more than three levels and limited work on categorical outcomes for CRTs with more than two levels. In particular, we know of only one approach for four-level stepped wedge CRTs with methodology based on a linear mixed model \citep{teerenstra2019}. Existing methods and their scope for designing trials with more than two levels are presented in Table \ref{tb:Tablesum1}. In particular, our two motivating examples pertain to CRTs with four levels, which necessitates new considerations on the within-cluster correlation structure and sample size determination. The first example is the Reducing Stigma among Healthcare Providers (RESHAPE) CRT conducted in Nepal, which evaluates a new intervention to improve accuracy of mental illness diagnosis in comparison to implementation as usual (IAU). In the RESHAPE CRT, binary diagnosis outcomes are observed for patients (level 1) nested in providers (level 2), who are nested in health facilities (level 3) within municipality (level 4), which is the unit of randomization. Figure \ref{fig:Fig1} provides a hierarchical illustration of the RESHAPE trial. Our second example is the Health and Literacy Intervention (HALI) trial conducted in Kenya, which compared a literacy intervention to improve early literacy outcomes to usual practice \citep{jukes2017}. In the HALI trial, repeated literacy continuous outcomes (level 1) are measured for children (level 2), who are nested within schools (level 3) of each Teacher Advisory Center (TAC) tutor zone (level 4).  The TAC tutor zone is the unit of randomization which, like the RESHAPE trial, is at the highest level and which we refer to as the \emph{cluster}.

\begin{table}[htbp]\footnotesize
\caption{Brief summary of existing methods for designing trials with more than two levels.}\label{tb:Tablesum1}
\centering
\begin{threeparttable}
\begin{tabular}{ccccc}
\toprule
%\midrule
Reference & Design & Model\tnote{a} & Outcome & Link \\
\midrule\smallskip
\citet{heo2008statistical} & Three-level CRT & GLMM & Continuous & Identity \\\midrule
\multirow{2}[3]{*}{\citet{teerenstra2010}} & \multirow{2}[3]{*}{Three-level CRT} & \multirow{2}[3]{*}{GEE} & Continuous & Identity \\\cmidrule{4-5}
& & & Binary & Logit \\\midrule
\citet{cunningham2016design} & Three-level Clustered Trial & GLMM & Continuous & Identity \\\midrule
& & & Continuous & Identity \\\cmidrule{4-5}
\citet{teerenstra2019} & Four-level stepped-wedge CRT & GLMM & Binary & Identity \\\cmidrule{4-5}
& & & Count & Identity \\\midrule
& & & Continuous & Identity\\\cmidrule{4-5}
\citet{liu2020} & Three-level CRT & GEE & Binary & Logit \\\cmidrule{4-5}
& & & Count & Log \\
\bottomrule
\end{tabular}\smallskip
\begin{tablenotes}\scriptsize
\item[a] GLMM: generalized linear mixed model. GEE: generalized estimating equations.
\end{tablenotes}
\end{threeparttable}
\end{table}

%%%%%%%%%%
% Figure 1
%%%%%%%%%%

To account for multiple levels of clustering, both marginal (population-averaged) and conditional (cluster-specific) models can be used to analyze the outcome data. In this article, we focus on developing new sample size procedures for four-level CRTs based on the marginal model due to its population-averaged interpretation which is of direct relevance to health-systems level research questions \citep{preisser2003}. In particular, the intervention effect parameter in the marginal model describes the difference of average outcome between the source population included in the intervention and control clusters, and its interpretation is unaffected by the specification of the within-cluster correlation structure \citep{zeger1988}. We characterize an \emph{extended nested exchangeable} correlation structure appropriate for the four-level CRT, and develop general sample size equations assuming arbitrary link and variance functions. Special cases of continuous, binary and count outcomes with commonly-used link functions are presented, which generalize existing sample size formulas for two-level and three-level CRTs \citep{shih1997,heo2008statistical,teerenstra2010}. When the randomization is carried out at the highest level, we show that the variance inflation factor (VIF) due to the four-level clustered structure is equal to the largest eigenvalue of the extended nested exchangeable correlation structure, which depends on the number of units at each level and on three intraclass correlation coefficients (ICCs). We carried out an extensive simulation study to demonstrate the accuracy and robustness of the proposed sample size formula with a binary outcome, and applied our method to design the RESHAPE and HALI trials. Finally, although we focus on randomization at the highest level as in our motivating CRTs, we also develop more general results for sample size calculation to allow randomization at lower levels.

\section{Generalized estimating equations and finite-sample adjustments}\label{sec:gee}
We first consider a four-level CRT, and use the RESHAPE trial (Figure \ref{fig:Fig1}) as a running example for illustration. Let $Y_\ii$ denote the outcome for patient $l = 1,\ldots, L_{ijk}$ from provider $k = 1,\ldots, K_{ij}$ nested in health facility $j = 1,\ldots, M_i$ of municipality $i = 1,\ldots, N$, and $\bfX_\ii = (X_{ijkl1},\ldots, X_{ijklp})'$ denote a list of $p$ covariates. Frequently in designing CRTs, $\bfX_\ii$ only includes an intercept and a cluster-level intervention indicator, \textcolor{black}{which equals one if the cluster is assigned to intervention and zero if the cluster is assigned to control.} Let $\mu_\ii = \text{E}(Y_\ii|\bfX_\ii)$ be the marginal mean outcome given $\bfX_\ii$, which is specified via a generalized linear model
\begin{align}\label{eq:mean}
    g(\mu_\ii) = \bfX_\ii'\bfbeta,
\end{align}
where $g$ is a link function and $\bfbeta$ is a $p\times 1$ vector of regression parameters. The marginal variance function is specified as var$(Y_\ii|\bfX_\ii) =\phi\nu_{ijkl}$, where $\phi$ is the common dispersion parameter and \textcolor{black}{$\nu_\ii$ is an arbitrary variance function that depends on the marginal mean and possibly additional dispersion parameters}. Denote $\kappa_\ii = \sqrt{\phi\nu_\ii}/\mu_\ii$ as the coefficient of variation (CV) of the outcome. In addition to the marginal mean model, we propose to characterize the degree of similarity among the within-cluster outcomes through an extended nested exchangeable correlation structure with the following assumptions: 
\begin{enumerate} 
\itemsep0em 
\item[(i)] the correlation between different patients from the same provider is \begin{equation*}
    \text{corr}(Y_\ii, Y_{ijkl'}|\bfX_\ii,\bfX_{ijkl'}) = \alpha_0,~~~~\text{for} ~l \neq l';
\end{equation*}
\item[(ii)] the correlation between patients from different providers but within the same health facility is 
\begin{equation*}
    \text{corr}(Y_\ii, Y_{ijk'l'}|\bfX_\ii,\bfX_{ijk'l'}) = \alpha_1,~~~~\text{for}~k \neq k';
\end{equation*}
\item[(iii)] the correlation between patients from different health facilities but within the same municipality is 
\begin{equation*}
    \text{corr}(Y_\ii, Y_{ij'k'l'}|\bfX_\ii, \bfX_{ij'k'l'}) = \alpha_2,~~~~\text{for}~j \neq j'.
\end{equation*}
\end{enumerate}
We term this correlation structure as the extended nested exchangeable structure because it accounts for an additional hierarchical nesting compared to the nested exchangeable structure developed for three-level CRTs in \citet{teerenstra2010}. Our three-correlation structure also differs from an existing three-parameter block exchangeable structure proposed for cohort stepped wedge designs \citep{li2018}. To illustrate these differences in correlation structures, we provide specific matrix examples in Web Appendix A.

For simplicity, we assume a balanced design so that there is an equal number of units at each level across all clusters, i.e., $M_i = M\geq 2\ (i = 1,\ldots, N)$, $K_{ij} = K\geq 2\ (i = 1,\ldots, N; j = 1,\ldots, M_i)$, and $L_{ijk} = L\geq 2\ (i = 1,\ldots, N; j = 1,\ldots, M_i; k = 1,\ldots, K_{ij})$. For each provider, let $\bfY_{ijk} = (Y_{ijk1},\ldots, Y_{ijkL})'$ and $\bfmu_{ijk} = (\mu_{ijk1},\ldots, \mu_{ijkL})'$ be the $L \times 1$ vector of outcomes and $L \times 1$ marginal mean vector, respectively. Furthermore, in each municipality, denote $\bfY_i = (\bfY_{i11}', \bfY_{i12}',\ldots, \bfY_{iMK}')'$, $\bfmu_i = (\bfmu_{i11}', \bfmu_{i12}', \ldots, \bfmu_{iMK}')'$, where $\bfY_i$ and $\bfmu_i$ are of dimension $MKL$ and include $MK$ blocks of provider-specific outcome vectors (classified by combinations of different health facilities and providers); denote $\bfX_i$ as the $MKL \times p$ covariate matrix. We use the generalized estimating equations (GEE) approach \citep{liang1986} to estimate the intervention effect in mean model \eqref{eq:mean}. Define $\bfD_i = \partial\bfmu_i/\partial\bfbeta'$, and let $\bfV_i = \bfA_i^{1/2}\bfR_i\bfA_i^{1/2}$ be a working covariance matrix for $\bfY_i$, where $\bfA_i$ is a $MKL$-dimensional diagonal matrix with elements of $\phi\nu_\ii$, and $\bfR_i = \bfR_i(\bfalpha)$ is a working correlation matrix specified by the ICC vector $\boldsymbol{\alpha} = (\alpha_0, \alpha_1, \alpha_2)'$. The extended nested exchangeable correlation structure can be concisely represented as
\begin{align}\label{eq:wc}
    \bfR_i = (1-\alpha_0)\bfI_{MKL} + (\alpha_0-\alpha_1)\bfI_{MK}\otimes \bfJ_L + (\alpha_1-\alpha_2)\bfI_M\otimes \bfJ_{KL} + \alpha_2\bfJ_{MKL},
\end{align}
where $\otimes$ is the Kronecker product, $\bfJ_u=\bfone_u\bfone_u'$ is a $u\times u$ matrix of ones, and $\bfI_u$ is a $u\times u$ identity matrix. Of note, when we equate $\alpha_0=\alpha_1$ or $\alpha_1=\alpha_2$, $\bfR_i$ reduces to the two-parameter nested exchangeable correlation structure developed in \citet{teerenstra2010}. The following Theorem provides a closed-form characterization of the eigenvalues of $\bfR_i$. The proof of Theorem \ref{thm:eigen} can be found in Web Appendix B.

\begin{theorem}\label{thm:eigen}
\textcolor{black}{The extended nested exchangeable correlation structure has the following eigenvalues}:
\begin{align*}
    &\lambda_1 = 1-\alpha_0,\\
    &\lambda_2 = 1 + (L-1)\alpha_0 - L\alpha_1,\\
    &\lambda_3 = 1 + (L-1)\alpha_0 +L(K-1)\alpha_1 - LK\alpha_2,\\
    &\lambda_4 = 1 + (L-1)\alpha_0 +L(K-1)\alpha_1 + LK(M-1)\alpha_2,
\end{align*}
with multiplicity $MK(L-1)$, $M(K-1)$, $M-1$ and $1$, respectively.
\end{theorem}

\begin{remark}
When (a) $\alpha_0 \neq \alpha_1$, $\alpha_1 \neq \alpha_2$, and $\alpha_2 \neq 0$, and (b) $\alpha_0 \neq - (K-1)\alpha_1 + K\alpha_2$, $\alpha_0 \neq -(K-1)\alpha_1 - K(M-1)\alpha_2$, and $\alpha_1 \neq -(M-1)\alpha_2$, Theorem \ref{thm:eigen} implies that the extended nested exchangeable correlation structure has four distinct eigenvalues. Otherwise, at least two elements of $\{\lambda_1,\lambda_2,\lambda_3,\lambda_4\}$ are identical and the multiplicity of each distinct eigenvalue can be inferred from Theorem \ref{thm:eigen} by simple addition. For example, if $\alpha_0 = \alpha_1 (= 0 \text{ or not})$, but the rest of the conditions in (a) and (b) hold, the extended nested exchangeable correlation structure has three distinct eigenvalues, with $\lambda_1 = \lambda_2$ whose multiplicity is $MK(L-1)+M(K-1)=M(KL-1)$.
\end{remark}

These explicit forms of the eigenvalues developed in Theorem \ref{thm:eigen} facilitate an efficient determination of the joint validity of the correlation parameters. Specifically, valid values for $(\alpha_0, \alpha_1, \alpha_2)$ should ensure a positive definite $\bfR_i$ and are contained in the convex open set defined by $\text{min}\{\lambda_1, \lambda_2, \lambda_3, \lambda_4\} > 0$. When the marginal mean model includes only an intercept and cluster-level intervention status, \textcolor{black}{there is often an additional natural restriction on the range of $\bfalpha$ for binary outcomes given in Equation (8) of \citet{qaqish2003}. When $\bfX_\ii$ only includes an intercept and a cluster-level intervention indicator, it is straightforward to show that the upper bound of $\bfalpha$ is one, and the lower bound is negative (this lower bound depends on $\text{P}_0$ and $\text{P}_1$, which are the marginal prevalences of the outcome in the control and intervention arms, respectively). In practice, the ICC values are assumed to be positive and the natural constraints are satisfied. We will maintain the positive ICC assumption for the rest of the article.}

In general, the GEE estimator $\hat\bfbeta$ solves the $\bfbeta$-estimating equations $\sum_{i=1}^{N} \bfD_i' \bfV_i^{-1}\left(\bfY_i-\bfmu_i\right)=\mathbf{0}$. Because the ICC parameters are of interest when analyzing CRTs, we specify a second set of $\bfalpha$-estimating equations to iteratively update the correlations that parameterize $\bfV_i$. In the presence of an unknown dispersion parameter, an additional estimating equation for $\phi$ needs to be specified beyond the $\bfbeta$-estimating equations and $\bfalpha$-estimating equations. To provide a correction to the small-sample bias in estimating $\bfalpha$ due to a limited number of clusters, we adopt the matrix-adjusted estimating equations (MAEE) of \citet{preisser2008}; also see Appendix B of \citet{li2018} for details of MAEE. As the number of clusters increases, $N^{1/2}(\hat\bfbeta-\bfbeta)$ converges to a multivariate normal random vector with mean $0$ and covariance estimated by the model-based variance $\bfV_{R}^{\text {model}}=\hat{\bfSigma}_{1}^{-1}=\left\{N^{-1}\sum_{i=1}^{N} \bfD_i'(\hat{\bfbeta})\bfV_i^{-1}(\hat{\bfalpha}) \bfD_i(\hat{\bfbeta})\right\}^{-1}$, or by the sandwich variance $\bfV_{R}^{\text {sandwich}}= \hat{\bfSigma}_{1}^{-1} \hat{\bfSigma}_{0} \hat{\bfSigma}_{1}^{-1}$ where
\begin{align}\label{eq:sw}
    \hat{\bfSigma}_{0}=& N^{-1}\sum_{i=1}^{N} \bfC_i\bfD_i'(\hat{\bfbeta}) \bfV_i^{-1}(\hat{\bfalpha}) (\bfY_i-\hat{\bfmu}_i)(\bfY_i-\hat{\bfmu}_i)'\bfV_i^{-1}(\hat{\bfalpha}) \bfD_i(\hat{\bfbeta})\bfC_i',
\end{align}
and $\bfC_i$ represents a multiplicative factor for small-sample bias correction in variance estimation. \textcolor{black}{While both $\bfV_{R}^{\text {model}}$ and $\bfV_{R}^{\text {sandwich}}$  provide adequate quantification of the uncertainty in estimating $\hat{\bfbeta}$ when the extended nested exchangeable correlation structure is correctly specified, only the sandwich variance $\bfV_{R}^{\text {sandwich}}$ is asymptotically valid when the working correlation matrix is misspecified. For example, the independence working correlation matrix is misspecified when the true correlation structure follows the extended nested exchangeable structure. Furthermore, the extended nested exchangeable structure can also be misspecified if the true correlation structure is more complex with more than three ICC parameters, for instance, when the ICC parameters depend on covariates such as intervention arm or other cluster characteristics. In this article, we consider only a single example of misspecification, namely the first case with the independence working correlation matrix, with details discussed in Section \ref{sec:rv}.}

In pragmatic CRTs, there is usually a limited number of units at the highest level. In the RESHAPE trial, it is only possible to randomize a total of 24 municipalities. Therefore, adjustments to the sandwich variance $\bfV_{R}^{\text {sandwich}}$ are required to reduce its potentially negative bias \citep{li2015}. Setting $\bfC_i = \bfI_p$ in \eqref{eq:sw} provides the uncorrected sandwich estimator of \citet{liang1986}, denoted as BC0. Because BC0 tends to underestimate the variance when the number of clusters is small, we consider four types of small-sample adjustments. Define matrix $\bfQ_i = \bfD_i'\bfV_i^{-1}\bfD_i(N\bfSigma_1)^{-1}$. Setting $\bfC_i = (\bfI_p-\bfQ_i)^{-1/2}$ provides the bias-corrected variance of \citet{kauermann2001}, or BC1 (here we provide an equivalent representation based on the $\bfQ_i$ matrix instead of the cluster-leverage matrix; we provide additional details in Web Appendix C to clarify this subtle equivalence). Setting $\bfC_i = (\bfI_p-\bfQ_i)^{-1}$ provides the bias-corrected variance of \citet{mancl2001}, or BC2. Setting $\bfC_i=\operatorname{diag}\left[\{1-\min (r,[\bfQ_i]_{j j})\}^{-1/2}\right]$, where $r < 1$ is a user-defined bound with a default value of 0.75, provides the bias-corrected variance of \citet{fay2001}, or BC3. Based on the degree of multiplicative adjustments, we generally have BC0 $<$ BC1 $\approx$ BC3 $<$ BC2 \citep{preisser2008,li2018}. In addition to the three multiplicative adjustments, we also consider an additive bias correction of \citet{morel2003}, defined as BC4 = $c\text{BC0} + \delta_N\phi\bfSigma_1^{-1}$, where $c = \{(f-1)/(f-p)\}\times \{N/(N-1)\}$, $f = \sum_{i=1}^{N}\sum_{j=1}^{M}\sum_{k=1}^{K}L_{ijk}$ is the total number of observations, $\delta_N = \text{min}\{0.5, p/(N-p)\}$ is the correction factor converging to 0 as $N$ tends to infinity, and 
\begin{equation*}
    \phi = \text{max}\left[1, \text{trace}\left\{c\left(\sum_{i=1}^N\bfD_i'\bfV_i^{-1}(\bfY_i-\hat{\bfmu}_i)(\bfY_i-\hat{\bfmu}_i)'\bfV_i^{-1}\bfD_i\right)(N\bfSigma_1)^{-1}\right\}/p\right].
\end{equation*}
One potential advantage of the additive bias correction BC4 over the multiplicative bias correction is that the former guarantees the positive definiteness of the estimated covariance \citep{morel2003}. The general MAEE methods are implemented in the recent R package \texttt{geeCRT} \citep{geeCRT}. Source R code to implement MAEE in four-level CRTs (including BC4) are also available  at
\break \url{https://github.com/XueqiWang/Four-Level_CRT}.

\section{Power and sample size considerations}\label{sec:power}
\textcolor{black}{At the design stage of a balanced four-level clustered trial, we consider an unadjusted marginal model \eqref{eq:mean} with $\bfX_{ijkl}$ including only an intercept and a binary cluster-level intervention indicator ($p=2$). This simple structure is assumed for the rest of the article.} Suppose we are interested in testing the null hypothesis of no intervention effect $H_0: \beta_2 = 0$, using a two-sided $t$-test. Specifically, the asymptotic distribution of $\sqrt{N}(\hat\beta_2 - \beta_2)$ is normal with mean $0$ and variance obtained as the lower-right element of cov$\{\sqrt{N}(\hat\bfbeta-\bfbeta)\}$. The Wald test statistic $\sqrt{N}\hat\beta_2/\sigma_\beta$, where $\sigma_\beta^2 = \text{var}(\sqrt{N}\hat\beta_2)$, will be compared to a $t$-distribution with $N-2$ degrees of freedom; the $t$-test is chosen because it often improves the test size in finite samples compared to normal approximations \citep{li2015, teerenstra2010}. The predicted power to detect an effect size $b$ on the link function scale with a nominal type I error rate $\alpha$ is then
\begin{align}\label{eq:power}
    1-\gamma=\Phi_{t, N-2}\left(t_{\alpha / 2, N-2} + \mathopen|b\mathopen|\sqrt{\frac{N}{\sigma_{\beta}^2}}\right),
\end{align}
where $\Phi_{t, d}$ and $t_{\alpha, d}$ are the cumulative distribution function and $100 \alpha \%$ percentile of the $t$-distribution with $d$ degrees of freedom, respectively. Accordingly, the number of clusters or level-four units required to achieve $100(1-\gamma)\%$ power must satisfy
\begin{align}\label{eq:ss}
    N\geq \left(t_{\alpha / 2, N-2}+t_{\gamma, N-2}\right)^{2} \frac{\sigma_{\beta}^{2}}{b^{2}}.
\end{align}
\textcolor{black}{Note that the above two equations are approximations, and Equation \eqref{eq:ss} must be solved iteratively.} Equations \eqref{eq:power} and \eqref{eq:ss} suggest that analytical power and sample size calculations depend on an explicit expression for the asymptotic variance $\sigma_\beta^2$. In what follows, we will assume the true correlation matrix of $\bfY_i$ is extended nested exchangeable defined in Equation \eqref{eq:wc}, and determine the explicit form of $\sigma_\beta^2$ when the working correlation is the true correlation structure or when the working correlation assumes independence.

\subsection{A general sample size expression with randomization at level 4}\label{sec:general}
Mimicking our motivating applications, we consider a total of $N$ clusters or level-four units are randomized with \textcolor{black}{$\pi_c N$} clusters assigned to the control arm and $(1-\pi_c)N$ clusters to the intervention arm. Let $\bfX_c$ and $\bfX_t$ be the design matrix for clusters assigned to the control and intervention conditions, respectively, i.e., $\bfX_c=(\bfone_{MKL}, \bfzero_{MKL})$ and $\bfX_t=(\bfone_{MKL}, \bfone_{MKL})$. \textcolor{black}{Below, we consider a general sample size expression for outcomes with an arbitrary mean-variance relationship.} To do so, we further define $\mu_c$ and $\mu_t$ to be the marginal mean for the control and intervention arms, respectively. \textcolor{black}{Because the randomization is carried out at level 4, the marginal means of all observations in each particular cluster ($\mu$) are either all $\mu_c$ or all $\mu_t$, depending on the intervention assignment.} We also define $\kappa_c$ and $\kappa_t$ as the marginal CV of outcome in the control and intervention arms, respectively. Assuming an arbitrary link function $g$, we have \textcolor{black}{$\bfD_i = \left(\partial g(\mu_c)/\partial\mu_c\right)^{-1}\bfX_c$} and $\bfV_i = \mu_c^2\kappa_c^2\bfR_i$ for all $i$ in the control arm, \textcolor{black}{$\bfD_i = \left(\partial g(\mu_t)/\partial\mu_t\right)^{-1}\bfX_t$} and $\bfV_i = \mu_t^2\kappa_t^2\bfR_i$ for all $i$ in the intervention arm.

In this Section, we assume that the working correlation model is the extended exchangeable correlation structure, with correlation parameters estimated via the MAEE approach. Therefore $\sigma_\beta^2$ is the lower-right element of the model-based variance, given by the inverse of
\begin{align*}
    \bfSigma_1 &= N^{-1}\sum_{i=1}^{N}\bfD_i'\bfV_i^{-1}\bfD_i =\left(\bfone' \bfR^{-1}\bfone\right)\left(\begin{matrix}\pi_c \rho_c^{-2}+(1-\pi_c)\rho_t^{-2} & (1-\pi_c)\rho_t^{-2} \\ (1-\pi_c)\rho_t^{-2} & (1-\pi_c)\rho_t^{-2} \end{matrix}\right),
\end{align*}
where $\bfone'\bfR^{-1}\bfone$ is the sum of all elements of the inverse correlation matrix $\bfR^{-1}$, $\rho_c = \mu_c\kappa_c \{\partial g(\mu_c)/\partial\mu_c\}$, and $\rho_t = \mu_t\kappa_t \{\partial g(\mu_t)/\partial\mu_t\}$. In Web Appendix D, we show that an explicit inverse of the extended nested exchangeable correlation matrix is
\begin{align*}
\bfR^{-1}=&\frac{1}{\lambda_1}\bfI_{MKL}
-\frac{\alpha_0-\alpha_1}{\lambda_1\lambda_2}\bfI_{MK}\otimes \bfJ_L
-\frac{\alpha_1-\alpha_2}{\lambda_2\lambda_3}\bfI_M\otimes \bfJ_{KL}-\frac{\alpha_2}{\lambda_3\lambda_4}\bfJ_{MKL},
\end{align*}
which gives $\bfone'\bfR^{-1}\bfone=MKL/\lambda_4$, where $\lambda_1$, $\lambda_2$, $\lambda_3$ and $\lambda_4$ are defined in Theorem \ref{thm:eigen}. These intermediate results allow us to obtain the asymptotic variance expression as
\begin{align}\label{eq:varbeta}
   \sigma_{\beta}^{2} = \frac{\lambda_4}{MKL}\left(\frac{\rho_c^2}{\pi_c} + \frac{\rho_t^2}{1-\pi_c}\right).
\end{align}
Hence the required number of clusters must satisfy
\begin{align}\label{eq:ssgen4}
    N\geq &\frac{\left(t_{\alpha / 2, N-2}+t_{\gamma, N-2}\right)^{2}}{b^{2}}\times \frac{\lambda_4}{MKL}\left(\frac{\rho_c^2}{\pi_c} + \frac{\rho_t^2}{1-\pi_c}\right).
\end{align}

The variance expression \eqref{eq:varbeta} has several important implications for study planning, which we discuss in a series of remarks below.

\begin{remark}
In the absence of any clustering, one can easily derive the asymptotic variance of the intervention effect estimator in an individually randomized trial as \begin{equation*}
    \tilde{\sigma}_\beta^2=\frac{1}{MKL}\left(\frac{\rho_c^2}{\pi_c} + \frac{\rho_t^2}{1-\pi_c}\right),
\end{equation*}
assuming a total of $N\times MKL$ individuals are recruited. As a consequence, the eigenvalue $\lambda_4=\sigma_\beta^2/\tilde{\sigma}_\beta^2$ is the design effect or VIF due to the four-level clustered structure. This VIF generalizes the previous findings in two-level and three-level CRTs, where the corresponding VIFs have also been found to be the largest eigenvalues of the simple exchangeable and nested exchangeable correlation matrices, respectively (see \citet{shih1997}, \citet{liu2019optimal} and discussions in Section 7 of \citet{li2019power} for VIF in two-level and three-level CRTs). 
\end{remark}

\begin{remark}
Because $\lambda_4=1 + (L-1)\alpha_0 +L(K-1)\alpha_1 + LK(M-1)\alpha_2$ is monotonically increasing in all three ICC parameters, $\alpha_0$, $\alpha_1$ and $\alpha_2$, larger values of any of these three ICC values would inflate the variance and decrease the design efficiency. Interestingly, the coefficient of each ICC parameter in the equation for $\lambda_4$ determines the relative importance of that ICC for variance inflation, and the sample size or variance inflation is most sensitive to changes in the correlation between patients from different health facilities within the same municipality (or the ICC at the highest hierarchy), $\alpha_2$, which has the largest coefficient $LK(M-1)$. On the other hand, ignoring the fourth level of clustering by assuming $\alpha_1=\alpha_2$ (as in a three-level CRT) could lead to a highly inaccurate power calculation if these two correlations are, in fact, different. That is, assuming the value of $\alpha_1$ is fixed and known, when the true value of $\alpha_2$ is larger than $\alpha_1$, the required sample size would be underestimated; when the true value of $\alpha_2$ is smaller than $\alpha_1$, the required sample size would be overestimated. Ignoring the ICC at the lowest hierarchy, $\alpha_0$, however, may have a relatively smaller impact on the power calculation, especially when the patient panel size $L$ is not too large.
\end{remark}

\begin{remark}\label{rmk:opt}
Expression \eqref{eq:varbeta} also suggests the optimal allocation of clusters does not depend on the ICC parameters. Specifically, the optimal randomization proportion that leads to the smallest $\sigma_\beta^2$ is obtained as $\pi_c^{\text{opt}}=1/2$ if $\rho_c=\rho_t$, and $\pi_c^{\text{opt}}=(\rho_c^2 \pm \mathopen|\rho_c\rho_t\mathopen|)/(\rho_c^2-\rho_t^2)$ (depending on which value is contained within the unit interval) otherwise. \textcolor{black}{Details of the derivation for $\pi_c^{\text{opt}}$ are provided in Web Appendix E.}
\end{remark}

\textcolor{black}{Finally, for commonly-used link functions and different types of outcomes, we can elegantly specify the marginal variance $\nu$ as a function of the marginal mean $\mu$. In those cases, sample size Equation \eqref{eq:ssgen4} can be recognized as more familiar expressions, examples of which are provided below.}

\begin{example}
(\textbf{Sample size formula with a continuous outcome}) When $Y_\ii$ is continuous, and a Gaussian variance function is assumed, we have \textcolor{black}{$\nu = 1$} for both trial arms, and the marginal CV of the outcome $\kappa_c=\sqrt{\phi}/\mu_c$ and $\kappa_t=\sqrt{\phi}/\mu_t$. Under the identity link function, $\partial g(\mu_t)/\partial\mu_t=\partial g(\mu_c)/\partial\mu_c=1$, and Equation \eqref{eq:ssgen4} reduces to 
\begin{align}\label{eq:sscon}
    N\geq &\frac{\left(t_{\alpha / 2, N-2}+t_{\gamma, N-2}\right)^{2}}{b^{2}}\times \frac{\phi\lambda_4}{\pi_c(1-\pi_c)MKL}.
\end{align}
In this case, $\phi$ is assumed as the homoscedastic marginal total variance of $Y_{ijkl}$, and the sample size expression \eqref{eq:sscon} extends the results of \citet{heo2008statistical} and \citet{teerenstra2010} to account for an additional level of clustering. Additionally, the optimal randomization proportion $\pi_c^{\text{opt}}$ is clearly $1/2$ given $\rho_c=\rho_t$. 
\end{example}

\begin{example}
(\textbf{Sample size formulas with a binary outcome}) When $Y_\ii$ is binary, the dispersion $\phi=1$, and the variance function $\nu = \mu(1-\mu)$. This leads to the CV of the outcome in the control and intervention arms as $\kappa_c=\sqrt{(1-\mu_c)/\mu_c}$ and $\kappa_t=\sqrt{(1-\mu_t)/\mu_t}$. We consider the sample size formulas for three commonly-used link functions. Under the canonical logit link function, $g(\mu) = \text{log}\left(\mu/(1-\mu)\right)$, with $\partial g(\mu)/\partial\mu= \mu^{-1}(1-\mu)^{-1}$. Define $\text{P}_0 = \left(1+\text{exp}(-\beta_1)\right)^{-1}$, $\text{P}_1 = \left(1+\text{exp}(-\beta_1-\beta_2)\right)^{-1}$ as the prevalence in the two arms, the required number of clusters to detect an effect size on the log odds ratio scale $b = \text{log}(\text{P}_1/(1-\text{P}_1))-\text{log}(\text{P}_0/(1-\text{P}_0))$ must satisfy
\begin{align}\label{eq:ssbinOR}
    N\geq &\frac{\left(t_{\alpha / 2, N-2}+t_{\gamma, N-2}\right)^{2}}{b^{2}}\times \frac{\lambda_4}{MKL}\left\{\frac{1}{\pi_c \text{P}_0\left(1-\text{P}_0\right)}+\frac{1}{(1-\pi_c) \text{P}_1\left(1-\text{P}_1\right)}\right\}.
\end{align}
Similarly, with an identity link function, we define the prevalence of outcome in the control and intervention clusters by $\text{P}_0 = \beta_1$, $\text{P}_1 = \beta_1 + \beta_2$, which should satisfy the boundary constraints $\text{P}_0,\text{P}_1\in(0,1)$. The general Equation \eqref{eq:ssgen4} suggests that the required number of clusters to detect an effect size on the risk difference scale $b=\text{P}_1-\text{P}_0$ satisfies
\begin{align}\label{eq:ssbinRD}
    N\geq &\frac{\left(t_{\alpha / 2, N-2}+t_{\gamma, N-2}\right)^{2}}{b^{2}}\times \frac{\lambda_4}{MKL}\left\{\frac{\text{P}_0\left(1-\text{P}_0\right)}{\pi_c}+\frac{\text{P}_1\left(1-\text{P}_1\right)}{1-\pi_c}\right\}.
\end{align}
Finally, under the log link function, we write $\text{P}_0 = \text{exp}(\beta_1)$ and $\text{P}_1 = \text{exp}(\beta_1+\beta_2)$, and define the effect size on the log relative risk scale to be $b = \text{log}(\text{P}_1) - \text{log}(\text{P}_0)$. The required sample size is provided by 
\begin{align}\label{eq:ssbinRR}
    N\geq &\frac{\left(t_{\alpha / 2, N-2}+t_{\gamma, N-2}\right)^{2}}{b^{2}}\times \frac{\lambda_4}{MKL}\left\{\frac{1-\text{P}_0}{\pi_c \text{P}_0}+\frac{1-\text{P}_1}{(1-\pi_c)\text{P}_1}\right\}.
\end{align}
Equations \eqref{eq:ssbinOR}, \eqref{eq:ssbinRD} and \eqref{eq:ssbinRR} generalize the sample size requirements provided in \citet{teerenstra2010} and \citet{liu2019optimal} from three-level CRTs to four-level CRTs.
\end{example}

\begin{example}
(\textbf{Sample size formula with a count outcome})
When $Y_\ii$ is a count outcome, a typical approach is to assume $Y_{\ii}$ follows the Poisson distribution. This assumption leads to $\phi=1$, $\nu = \mu$, and the CV of the Poisson outcome in each arm is $\kappa_c=\sqrt{1/\mu_c}$, $\kappa_t=\sqrt{1/\mu_t}$. Under the canonical log link function, $\partial g(\mu)/\partial\mu= 1/\mu$. Therefore, we define $\mu_c = \text{exp}(\beta_1)$ and $\mu_t = \text{exp}(\beta_1+\beta_2)$ as the expected event rate in the two arms, and the required number of clusters to detect an effect size on the log rate ratio scale $b=\log(\mu_t/\mu_c)$ is given by
\begin{align}\label{eq:sscount_poi}
    N\geq &\frac{\left(t_{\alpha / 2, N-2}+t_{\gamma, N-2}\right)^{2}}{b^{2}}\times \frac{\lambda_4}{MKL}\left\{\frac{1}{\pi_c \text{exp}(\beta_1)}+\frac{1}{(1-\pi_c) \text{exp}(\beta_1+\beta_2)}\right\}.
\end{align}
The sample size equation \eqref{eq:sscount_poi} generalizes the result in \citet{amatya2013sample} from two-level CRTs to four-level CRTs. 
\end{example}

\subsection{Randomization at lower levels}\label{sec:ol}
Although our two motivating examples represent CRTs that randomize at the fourth level, there may sometimes be administrative or practical considerations for randomizing units at a lower level. For example, when such lower-level randomization would not lead to intervention contamination, there would be an increase in the number of units of randomization and hence in the statistical power. Another intuition for improved power is that the inclusion of within-cluster contrasts in the intervention effect contributes to the increase in effective sample size. For completeness, we provide some general results for randomizing at lower levels. In our running example, the RESHAPE trial, randomizing at level 3 refers to, within each of the $N$ municipalities, randomizing $\pi_c M$ health facilities to the control arm and $(1-\pi_c)M$ health facilities to the intervention arm. Likewise, randomizing at level 2 refers to, within each of $NM$ health facilities, randomizing $\pi_c K$ providers in the control arm and $(1-\pi_c)K$ providers in the intervention arm. Because patients are nested within each provider, a design that randomizes at level 3 or level 2 within the four-level clustered structure would still be classified as a CRT. In contrast, randomizing at level 1 would be an individually randomized trial for which  $\pi_c L$ patients and $(1-\pi_c)L$ patients are randomized to the control and intervention arms, respectively, within each of the $NMK$ providers. Notice that the design vector $\bfX_{ijkl}$ can depend on the level of randomization, and we provide full details in Web Appendix F. To link these different contexts, we establish the following Theorem that applies to randomization at any level. The proof is in Web Appendix F.

\begin{theorem}\label{thm:ss}
Consider a four-level clustered trial, suppose the randomization is carried out at the $r$th level ($r=1,2,3,4$), and assume arbitrary link and variance functions. The asymptotic model-based variance for the GEE estimator $\hat{\beta}_2$ that uses the extended nested exchangeable working correlation is
\begin{align}\label{eq:ssgen}
    \sigma_{\beta}^{2} = \frac{\lambda_r}{MKL}\left(\frac{\rho_c^2}{\pi_c} + \frac{\rho_t^2}{1-\pi_c}\right)+\frac{(\lambda_4-\lambda_r)(\rho_c-\rho_t)^2}{MKL},
\end{align}
where $\rho_c = \mu_c\kappa_c \{\partial g(\mu_c)/\partial\mu_c\}$ and $\rho_t = \mu_t\kappa_t \{\partial g(\mu_t)/\partial\mu_t\}$, $\mu_c$ and $\mu_t$ are the marginal means for the control and intervention population, respectively, $\kappa_c$ and $\kappa_t$ are the marginal coefficients of variation of the outcome for the control and intervention population, respectively.
\end{theorem}

Clearly, Equation \eqref{eq:ssgen} includes Equation \eqref{eq:varbeta} as a special case, where the second term of \eqref{eq:ssgen} vanishes. Furthermore, Theorem \ref{thm:ss} has several interesting implications, which we elaborate below in several remarks. 

\begin{remark}
When randomizing at a lower level $r\leq 3$, the design effect relative to an individually randomized trial without any clustering is given by
\begin{align}\label{eq:de}
\text{design effect}=\lambda_r+(\lambda_4-\lambda_r)(\rho_c-\rho_t)^2
\left(\frac{\rho_c^2}{\pi_c} + \frac{\rho_t^2}{1-\pi_c}\right)^{-1},
\end{align}
which generally depends on the distribution of the outcome only through the first two moments via $\rho_c$ and $\rho_t$. With a continuous outcome, identity link function and a Gaussian variance function, we have $\rho_c=\rho_t$, and the design effect \eqref{eq:de} when randomizing at level $r$ simply reduces to the corresponding eigenvalue $\lambda_r$. 
This result indicates that as we randomize at a lower level, the design efficiency depends on the relative magnitude of the ICC parameters, because $\lambda_2=\lambda_1+L(\alpha_0-\alpha_1)$, 
$\lambda_3=\lambda_2+LK(\alpha_1-\alpha_2)$, and 
$\lambda_4=\lambda_3+LKM\alpha_2$. In our HALI trial context where we expect $\alpha_0\geq \alpha_1\geq \alpha_2$, randomization at a lower level leads to higher efficiency with a continuous outcome. These insights with a continuous outcome are in parallel to the design effect expressions provided in \citet{cunningham2016design} in a three-level design with a linear mixed model. 
\end{remark}

\begin{remark}
Expression \eqref{eq:ssgen} suggests that the optimal allocation proportion does not change according to the level of randomization. It is straightforward to see from Equation \eqref{eq:ssgen} that the optimal randomization probability $\pi_c^{\text{opt}}$ minimizes $\{\rho_c^2/\pi_c+\rho_t^2/(1-\pi_c)\}$ and equals to the result derived in Remark \ref{rmk:opt}.
\end{remark}

Finally, specific sample size formulas for randomization at lower levels can be derived based on Equation \eqref{eq:ss} and \eqref{eq:ssgen}, under commonly-used link functions and variance functions as in Section \ref{sec:general}. Details of these specific cases are provided in Web Appendix G.

\subsection{Considerations on using an independence working correlation}\label{sec:rv}
In our derivations of the variance expression $\sigma_\beta^2$, we have assumed that the working correlation structure is correctly specified as the extended nested exchangeable structure, and the ICCs are estimated through MAEE. In the context of clustered trials, using MAEE has been recommended because reporting accurate ICC estimates can inform the design of future trials \citep{preisser2008,teerenstra2010}, and it is considered good practice per the CONSORT extension to CRTs \citep{campbell2012}. Alternatively, one may pre-specify the primary analysis to be the GEE analysis with an independence working correlation and account for clustering simply via the sandwich variance. In a two-level design with equal cluster sizes, previous studies \citep{pan2001,yu2020evaluation,li2021truncation} have found that GEE estimators with working exchangeable and working independence have the same asymptotic efficiency when the randomization is carried out at the cluster level (i.e. the highest level in a two-level design). In Web Appendix H, we prove the following Theorem, which is a more general result that the same asymptotic equivalence holds in a four-level clustered design with equal ``cluster" sizes at each level, regardless of the level of randomization. 

\begin{theorem}\label{prop1}
Under balanced designs, suppose the randomization is carried out at the $r$th level ($r=1,2,3,4$), and assume arbitrary link and variance functions, with the extended nested exchangeable correlation structure as the true correlation structure. Using the sandwich variance with an independence working correlation matrix results in the same $\sigma^2_\beta$ as we obtain using the extended nested exchangeable working correlation in Theorem \ref{thm:ss}.
\end{theorem}

Therefore, Theorem \ref{thm:ss} still holds for GEE analysis assuming working independence and there is no difference in the derived sample size equations assuming equal ``cluster" sizes at each level, when randomization is carried out at any one of the four levels.

\section{Simulation study}\label{sec:sim}
Since we expect that the closed-form sample size formula for binary outcomes would be less accurate compared to those for continuous and count outcomes, we conducted a simulation study to evaluate the validity of our sample size formula for binary outcomes under the canonical logit link function for both balanced and unbalanced designs, where the latter has variable numbers of level-one units per level-two unit. Because we derived our sample size formulas assuming an equal number of patients within each provider, we consider the unbalanced design to assess the robustness of our formulas when the balance design assumption fails to hold in a specific and meaningful way (i.e., variable numbers of patients per provider in the RESHAPE trial). We focus on randomization at the fourth level with equal allocation to the two trial arms ($\pi_c=1/2$), which mimics the motivating studies. Correlated binary data in each cluster were generated from a binomial model with marginal mean in \eqref{eq:mean} and extended nested exchangeable correlation structure using the methods of \citet{qaqish2003}.

\subsection{Balanced designs}\label{sec:ecs}
We varied correlation values by choosing $\bfalpha = (\alpha_0, \alpha_1, \alpha_2)$ = \{(0.4, 0.1, 0.03), (0.15, 0.08, 0.02), (0.1, 0.02, 0.01), (0.05, 0.05, 0.02)\}; these values resemble estimates from the HALI trial \citep{jukes2017} or assumptions for the RESHAPE trial. Marginal mean parameters $\beta_1$ and $\beta_2$ were induced from the marginal means $\text{P}_0$ in the control arm and $\text{P}_1$ in the intervention arm for assessing the empirical power; $\beta_2$ was fixed at $0$ for assessing the empirical type I error rate. The nominal type I error rate was fixed at $5\%$. For illustration, we fixed $M\in\{2, 3\}$, $K \in\{ 3, 4\}$ and $L \in\{5, 10\}$. The total number of clusters $N$ was determined as the smallest number ensuring that the predicted power was $80\%$, and ranged from $8$ to $30$ across $30$ simulation scenarios. For each scenario, $1000$ data replications were generated and analyzed using GEE for the mean model and MAEE for the extended nested exchangeable working correlation structure. We consider $7$ variance estimators for the intervention effect: the model-based variance (MB), BC0, BC1, BC2, the variance estimator of \citet{ford2017improved} with standard error obtained as the average of those from BC1 and BC2 (denoted as AVG), BC3, and BC4. The convergence rate exceeded $99\%$ for most scenarios except for a few cases (a summary of convergence rates for all scenarios under balanced designs is presented in Web Table 1). \textcolor{black}{Since the nominal type I error rate was $5\%$, according to the margin of error from a binomial model with $1000$ replications, we considered an empirical type I error rate from $3.6\%$ to $6.4\%$ as acceptable. Similarly, since the predicted power was at least $80\%$ for each scenario, we considered an empirical power differing at most $2.6\%$ from the predicted power as acceptable. These acceptable bounds will be labeled with gray dashed lines in all related Figures summarizing the simulation results.}

Table \ref{tb:Table1} summarizes the results for empirical type I error rates for $t$-tests using different variance estimators. The type I error rates with BC1 were valid across almost all scenarios, except for only one scenario where the test became slightly liberal ($0.068$). BC0 often gave inflated type I error rates, while MB, BC2, AVG, BC3, and BC4 sometimes led to overly conservative type I error rates. Table \ref{tb:Table2} summarizes the results for predicted power and empirical power using different variance estimators. The empirical power with BC1 corresponded well with the predicted power throughout. BC0 provided higher empirical power than predicted in most scenarios, and MB sometimes led to higher empirical power than prediction; AVG, BC3, and BC4 sometimes gave lower empirical power than predicted, while BC2 almost always led to lower empirical power than predicted. Overall, the $t$-test with BC1 performed best; the performance among AVG, BC3, and BC4 were very similar.

%%%%%%%%%%
% Table 2
%%%%%%%%%%
\begin{table}[htbp]
\caption{Simulation scenarios and empirical type I error rates\textsuperscript{a} of GEE/MAEE analyses based on different variance estimators, using the extended nested exchangeable working correlation structure under balanced four-level CRTs. MB: Model-based variance estimator; BC0: Uncorrected sandwich estimator of \citet{liang1986}; BC1: Bias-corrected sandwich estimator of \citet{kauermann2001}; BC2: Bias-corrected sandwich estimator of \citet{mancl2001}; AVG: Bias-corrected sandwich estimator with standard error as the average of those from BC1 and BC2; BC3: Bias-corrected sandwich estimator of \citet{fay2001}; BC4: Bias-corrected sandwich estimator of \citet{morel2003}.}\label{tb:Table1}
\centering
\begin{threeparttable}
\begin{tabular}{lccrrrrccccccc}
\toprule
\midrule
%$\text{P}_0$ & $\text{P}_1$ & $\bfalpha$\tnote{b} & $n$ & $m$ & $k$ & $l$ & MB\tnote{c} & BC0\tnote{d} & BC1\tnote{e} & BC2\tnote{f} & AVG\tnote{g} & BC3\tnote{h} & BC4\tnote{i}\\
$\text{P}_0$ & $\text{P}_1$ & $\bfalpha$\tnote{b} & $n$ & $m$ & $k$ & $l$ & MB & BC0 & BC1 & BC2 & AVG & BC3 & BC4\\
\midrule
0.2 & 0.5 & A1 & 14 & 2 & 3 & 5 & \bftab 0.045 & 0.068 & \bftab 0.047 & \bftab 0.042 & \bftab 0.045 & \bftab 0.045 & \bftab 0.043 \\
0.2 & 0.5 & A1 & 14 & 2 & 3 & 10 & \bftab 0.046 & 0.078 & \bftab 0.056 & \bftab 0.041 & \bftab 0.047 & \bftab 0.049 & \bftab 0.046 \\
0.2 & 0.5 & A1 & 14 & 2 & 4 & 5 & \bftab 0.044 & 0.069 & \bftab 0.050 & \bftab 0.036 & \bftab 0.044 & \bftab 0.047 & \bftab 0.044 \\
0.2 & 0.5 & A1 & 12 & 3 & 3 & 5 & \bftab 0.046 & 0.070 & \bftab 0.049 & 0.033 & \bftab 0.042 & \bftab 0.038 & \bftab 0.039 \\
0.2 & 0.5 & A2 & 10 & 2 & 3 & 5 & 0.033 & \bftab 0.060 & \bftab 0.044 & 0.024 & 0.032 & 0.033 & 0.029 \\
0.2 & 0.5 & A2 & 10 & 2 & 3 & 10 & \bftab 0.045 & 0.077 & \bftab 0.054 & 0.032 & \bftab 0.041 & \bftab 0.040 & \bftab 0.037 \\
0.2 & 0.5 & A2 & 10 & 2 & 4 & 5 & \bftab 0.045 & 0.067 & \bftab 0.048 & 0.032 & \bftab 0.039 & \bftab 0.043 & \bftab 0.037 \\
0.2 & 0.5 & A2 & 8 & 3 & 3 & 5 & \bftab 0.036 & 0.071 & \bftab 0.038 & 0.021 & 0.029 & 0.030 & 0.028 \\
0.2 & 0.5 & A3 & 8 & 2 & 3 & 5 & 0.033 & \bftab 0.051 & 0.034 & 0.014 & 0.021 & 0.027 & 0.021 \\
0.2 & 0.5 & A3 & 8 & 3 & 3 & 5 & 0.031 & \bftab 0.060 & 0.030 & 0.016 & 0.021 & 0.023 & 0.020 \\ \smallskip
0.2 & 0.5 & A4 & 8 & 3 & 3 & 5 & 0.032 & \bftab 0.052 & 0.033 & 0.014 & 0.025 & 0.024 & 0.024 \\
0.1 & 0.3 & A1 & 22 & 2 & 3 & 5 & \bftab 0.037 & 0.070 & \bftab 0.055 & \bftab 0.047 & \bftab 0.048 & \bftab 0.050 & \bftab 0.047 \\
0.1 & 0.3 & A1 & 20 & 2 & 3 & 10 & \bftab 0.037 & 0.079 & 0.068 & \bftab 0.046 & \bftab 0.057 & \bftab 0.059 & \bftab 0.051 \\
0.1 & 0.3 & A1 & 20 & 2 & 4 & 5 & \bftab 0.039 & 0.073 & \bftab 0.055 & \bftab 0.045 & \bftab 0.049 & \bftab 0.050 & \bftab 0.046 \\
0.1 & 0.3 & A1 & 16 & 3 & 3 & 5 & 0.033 & 0.073 & \bftab 0.050 & \bftab 0.036 & \bftab 0.043 & \bftab 0.040 & \bftab 0.041 \\
0.1 & 0.3 & A2 & 16 & 2 & 3 & 5 & \bftab 0.038 & 0.074 & \bftab 0.055 & \bftab 0.040 & \bftab 0.046 & \bftab 0.047 & \bftab 0.044 \\
0.1 & 0.3 & A2 & 14 & 2 & 3 & 10 & \bftab 0.041 & 0.072 & \bftab 0.058 & \bftab 0.040 & \bftab 0.049 & \bftab 0.044 & \bftab 0.042 \\
0.1 & 0.3 & A2 & 14 & 2 & 4 & 5 & \bftab 0.042 & 0.076 & \bftab 0.051 & \bftab 0.037 & \bftab 0.047 & \bftab 0.044 & \bftab 0.040 \\
0.1 & 0.3 & A2 & 12 & 3 & 3 & 5 & \bftab 0.040 & 0.083 & \bftab 0.056 & \bftab 0.040 & \bftab 0.049 & \bftab 0.048 & \bftab 0.044 \\
0.1 & 0.3 & A3 & 12 & 2 & 3 & 5 & \bftab 0.042 & 0.070 & \bftab 0.057 & \bftab 0.041 & \bftab 0.048 & \bftab 0.046 & \bftab 0.045 \\
0.1 & 0.3 & A3 & 10 & 3 & 3 & 5 & \bftab 0.037 & \bftab 0.059 & \bftab 0.038 & 0.028 & 0.029 & 0.029 & 0.028 \\ \smallskip
0.1 & 0.3 & A4 & 10 & 3 & 3 & 5 & 0.033 & 0.066 & \bftab 0.042 & 0.025 & 0.034 & 0.032 & 0.029 \\
0.5 & 0.7 & A1 & 26 & 2 & 4 & 5 & \bftab 0.063 & 0.066 & \bftab 0.063 & \bftab 0.049 & \bftab 0.055 & \bftab 0.057 & \bftab 0.054 \\
0.5 & 0.7 & A2 & 16 & 3 & 3 & 5 & \bftab 0.059 & 0.065 & \bftab 0.057 & \bftab 0.041 & \bftab 0.049 & \bftab 0.049 & \bftab 0.049 \\
0.5 & 0.7 & A3 & 12 & 2 & 4 & 5 & \bftab 0.054 & 0.069 & \bftab 0.054 & \bftab 0.037 & \bftab 0.041 & \bftab 0.040 & \bftab 0.040 \\ \smallskip
0.5 & 0.7 & A4 & 14 & 3 & 3 & 5 & \bftab 0.048 & \bftab 0.061 & \bftab 0.039 & 0.033 & 0.035 & 0.035 & 0.034 \\
0.8 & 0.9 & A2 & 30 & 3 & 3 & 5 & \bftab 0.047 & \bftab 0.053 & \bftab 0.047 & \bftab 0.041 & \bftab 0.044 & \bftab 0.046 & \bftab 0.044 \\
0.8 & 0.9 & A3 & 22 & 2 & 4 & 5 & \bftab 0.046 & \bftab 0.056 & \bftab 0.047 & \bftab 0.037 & \bftab 0.039 & \bftab 0.039 & \bftab 0.040 \\
0.8 & 0.9 & A4 & 28 & 2 & 4 & 5 & \bftab 0.051 & \bftab 0.057 & \bftab 0.051 & \bftab 0.037 & \bftab 0.042 & \bftab 0.043 & \bftab 0.041 \\
0.8 & 0.9 & A4 & 24 & 3 & 3 & 5 & \bftab 0.050 & \bftab 0.056 & \bftab 0.049 & \bftab 0.041 & \bftab 0.046 & \bftab 0.046 & \bftab 0.046 \\
\bottomrule
\end{tabular}\smallskip
\begin{tablenotes}\small
\item[a] Bold text indicates acceptable empirical type I error rate (from $3.6\%$ to $6.4\%$).\smallskip
\item[b] A1: $\bfalpha=(0.4,0.1,0.03)$; A2: $\bfalpha=(0.15,0.08,0.02)$;
A3: $\bfalpha=(0.1,0.02,0.01)$; A4: $\bfalpha=(0.05,0.05,0.02)$.
%\smallskip
% \item[c] MB: Model-based variance estimator.\smallskip
% \item[d] BC0: Uncorrected sandwich estimator of Liang and Zeger (1986).\smallskip
% \item[e] BC1: Bias-corrected sandwich estimator of Kauermann and Carroll (2001).\smallskip
% \item[f] BC2: Bias-corrected sandwich estimator of Mancl and DeRouen (2001).\smallskip
% \item[g] AVG: Bias-corrected sandwich estimator with standard error as the average of those from BC1 and BC2.\smallskip
% \item[h] BC3: Bias-corrected sandwich estimator of Fay and Graubard (2001).\smallskip
% \item[i] BC4: Bias-corrected sandwich estimator of Morel, Bokossa, and Neerchal (2003).
\end{tablenotes}
\end{threeparttable}
\end{table}

%%%%%%%%%%
% Table 3
%%%%%%%%%%
\begin{table}[htbp]
\caption{Simulation scenarios, predicted power, and empirical power\textsuperscript{a} of GEE/MAEE analyses based on different variance estimators, using the extended nested exchangeable working correlation structure under balanced four-level CRTs. MB: Model-based variance estimator; BC0: Uncorrected sandwich estimator of \citet{liang1986}; BC1: Bias-corrected sandwich estimator of \citet{kauermann2001}; BC2: Bias-corrected sandwich estimator of \citet{mancl2001}; AVG: Bias-corrected sandwich estimator with standard error as the average of those from BC1 and BC2; BC3: Bias-corrected sandwich estimator of \citet{fay2001}; BC4: Bias-corrected sandwich estimator of \citet{morel2003}.}\label{tb:Table2}
\centering
\begin{threeparttable}
\begin{tabular}{lccrrrrcccccccc}
\toprule
\midrule
%$\text{P}_0$ & $\text{P}_1$ & $\bfalpha$\tnote{b} & $n$ & $m$ & $k$ & $l$ & Pred\tnote{c} & MB\tnote{d} & BC0\tnote{e} & BC1\tnote{f} & BC2\tnote{g} & AVG\tnote{h} & BC3\tnote{i} & BC4\tnote{j}\\
$\text{P}_0$ & $\text{P}_1$ & $\bfalpha$\tnote{b} & $n$ & $m$ & $k$ & $l$ & Pred\tnote{c} & MB & BC0 & BC1 & BC2 & AVG & BC3 & BC4\\
\midrule
0.2 & 0.5 & A1 & 14 & 2 & 3 & 5 & 0.817 & \bftab 0.821  & 0.857 & \bftab 0.822  & 0.767 & \bftab 0.798  & \bftab 0.795  & \bftab 0.797  \\
0.2 & 0.5 & A1 & 14 & 2 & 3 & 10 & 0.845 & \bftab 0.829  & 0.876 & \bftab 0.833  & 0.790 & 0.811 & 0.806 & 0.808 \\
0.2 & 0.5 & A1 & 14 & 2 & 4 & 5 & 0.866 & \bftab 0.891  & 0.909 & \bftab 0.884  & \bftab 0.854  & \bftab 0.871  & \bftab 0.869  & \bftab 0.870  \\
0.2 & 0.5 & A1 & 12 & 3 & 3 & 5 & 0.857 & \bftab 0.856  & 0.891 & \bftab 0.851  & 0.793 & 0.824 & 0.821 & 0.822 \\
0.2 & 0.5 & A2 & 10 & 2 & 3 & 5 & 0.808 & \bftab 0.814  & 0.881 & \bftab 0.811  & 0.743 & \bftab 0.782  & 0.776 & 0.776 \\
0.2 & 0.5 & A2 & 10 & 2 & 3 & 10 & 0.870 & \bftab 0.868  & 0.907 & \bftab 0.863  & 0.792 & 0.829 & 0.827 & 0.829 \\
0.2 & 0.5 & A2 & 10 & 2 & 4 & 5 & 0.852 & \bftab 0.860  & 0.897 & \bftab 0.853  & 0.784 & 0.823 & 0.820 & 0.820 \\
0.2 & 0.5 & A2 & 8 & 3 & 3 & 5 & 0.800 & \bftab 0.822  & 0.895 & \bftab 0.823  & 0.712 & \bftab 0.776  & 0.772 & 0.768 \\
0.2 & 0.5 & A3 & 8 & 2 & 3 & 5 & 0.851 & \bftab 0.859  & 0.916 & \bftab 0.848  & 0.752 & 0.793 & 0.794 & 0.785 \\
0.2 & 0.5 & A3 & 8 & 3 & 3 & 5 & 0.936 & \bftab 0.955  & 0.981 & \bftab 0.950  & 0.904 & \bftab 0.927  & \bftab 0.930  & \bftab 0.926  \\ \smallskip
0.2 & 0.5 & A4 & 8 & 3 & 3 & 5 & 0.892 & \bftab 0.914  & 0.957 & \bftab 0.907  & 0.830 & \bftab 0.869  & 0.865 & 0.860 \\
0.1 & 0.3 & A1 & 22 & 2 & 3 & 5 & 0.829 & 0.860 & 0.877 & \bftab 0.854  & \bftab 0.828  & \bftab 0.843  & \bftab 0.838  & \bftab 0.840  \\
0.1 & 0.3 & A1 & 20 & 2 & 3 & 10 & 0.818 & 0.849 & 0.865 & \bftab 0.841  & \bftab 0.813  & \bftab 0.830  & \bftab 0.827  & \bftab 0.829  \\
0.1 & 0.3 & A1 & 20 & 2 & 4 & 5 & 0.841 & 0.875 & 0.885 & \bftab 0.857  & \bftab 0.829  & \bftab 0.843  & \bftab 0.839  & \bftab 0.841  \\
0.1 & 0.3 & A1 & 16 & 3 & 3 & 5 & 0.805 & \bftab 0.831  & 0.851 & \bftab 0.817  & \bftab 0.783  & \bftab 0.804  & \bftab 0.803  & \bftab 0.805  \\
0.1 & 0.3 & A2 & 16 & 2 & 3 & 5 & 0.844 & 0.885 & 0.907 & 0.875 & \bftab 0.833  & \bftab 0.857  & \bftab 0.852  & \bftab 0.856  \\
0.1 & 0.3 & A2 & 14 & 2 & 3 & 10 & 0.849 & \bftab 0.862  & 0.894 & \bftab 0.854  & 0.807 & \bftab 0.827  & \bftab 0.826  & \bftab 0.829  \\
0.1 & 0.3 & A2 & 14 & 2 & 4 & 5 & 0.829 & 0.880 & 0.897 & 0.869 & \bftab 0.822  & \bftab 0.849  & \bftab 0.841  & \bftab 0.846  \\
0.1 & 0.3 & A2 & 12 & 3 & 3 & 5 & 0.826 & 0.858 & 0.892 & \bftab 0.844  & 0.798 & \bftab 0.828  & \bftab 0.825  & \bftab 0.825  \\
0.1 & 0.3 & A3 & 12 & 2 & 3 & 5 & 0.873 & 0.903 & 0.930 & \bftab 0.892  & \bftab 0.854  & \bftab 0.877  & \bftab 0.874  & \bftab 0.876  \\
0.1 & 0.3 & A3 & 10 & 3 & 3 & 5 & 0.898 & 0.928 & 0.959 & \bftab 0.916  & 0.871 & \bftab 0.892  & \bftab 0.890  & \bftab 0.891  \\ \smallskip
0.1 & 0.3 & A4 & 10 & 3 & 3 & 5 & 0.837 & 0.887 & 0.919 & 0.878 & \bftab 0.799  & \bftab 0.840  & \bftab 0.829  & \bftab 0.832  \\
0.5 & 0.7 & A1 & 26 & 2 & 4 & 5 & 0.823 & \bftab 0.841  & 0.854 & \bftab 0.838  & \bftab 0.814  & \bftab 0.825  & \bftab 0.825  & \bftab 0.824  \\
0.5 & 0.7 & A2 & 16 & 3 & 3 & 5 & 0.831 & \bftab 0.838  & 0.864 & \bftab 0.831  & 0.789 & \bftab 0.810  & \bftab 0.813  & \bftab 0.811  \\
0.5 & 0.7 & A3 & 12 & 2 & 4 & 5 & 0.827 & \bftab 0.829  & 0.860 & \bftab 0.825  & 0.778 & 0.800 & \bftab 0.806  & 0.797 \\ \smallskip
0.5 & 0.7 & A4 & 14 & 3 & 3 & 5 & 0.868 & \bftab 0.857  & \bftab 0.894  & \bftab 0.861  & 0.814 & 0.835 & 0.835 & 0.832 \\
0.8 & 0.9 & A2 & 30 & 3 & 3 & 5 & 0.804 & \bftab 0.825  & 0.838 & \bftab 0.818  & \bftab 0.800  & \bftab 0.810  & \bftab 0.810  & \bftab 0.811  \\
0.8 & 0.9 & A3 & 22 & 2 & 4 & 5 & 0.804 & \bftab 0.815  & \bftab 0.829  & \bftab 0.810  & \bftab 0.788  & \bftab 0.801  & \bftab 0.805  & \bftab 0.801  \\
0.8 & 0.9 & A4 & 28 & 2 & 4 & 5 & 0.824 & \bftab 0.844  & \bftab 0.850  & \bftab 0.839  & \bftab 0.826  & \bftab 0.836  & \bftab 0.837  & \bftab 0.836  \\
0.8 & 0.9 & A4 & 24 & 3 & 3 & 5 & 0.813 & \bftab 0.830  & 0.847 & \bftab 0.823  & \bftab 0.799  & \bftab 0.810  & \bftab 0.811  & \bftab 0.811 \\
\bottomrule
\end{tabular}\smallskip
\begin{tablenotes}\small
\item[a] Bold text indicates acceptable empirical power (differing at most $2.6\%$ from the predicted power).\smallskip
\item[b] A1: $\bfalpha=(0.4,0.1,0.03)$; A2: $\bfalpha=(0.15,0.08,0.02)$;
A3: $\bfalpha=(0.1,0.02,0.01)$; A4: $\bfalpha=(0.05,0.05,0.02)$.\smallskip
\item[c] Pred: Predicted power based on $t$-test.
% \smallskip
% \item[d] MB: Model-based variance estimator.\smallskip
% \item[e] BC0: Uncorrected sandwich estimator of Liang and Zeger (1986).\smallskip
% \item[f] BC1: Bias-corrected sandwich estimator of Kauermann and Carroll (2001).\smallskip
% \item[g] BC2: Bias-corrected sandwich estimator of Mancl and DeRouen (2001).\smallskip
% \item[h] AVG: Bias-corrected sandwich estimator with standard error as the average of those from BC1 and BC2.\smallskip
% \item[i] BC3: Bias-corrected sandwich estimator of Fay and Graubard (2001).\smallskip
% \item[j] BC4: Bias-corrected sandwich estimator of Morel, Bokossa, and Neerchal (2003).
\end{tablenotes}
\end{threeparttable}
\end{table}

The above simulations were repeated with data fit using GEE under working independence. The results for empirical type I error rates are summarized in Web Table 2, analogous to Table \ref{tb:Table1}; the results for power are summarized in Web Table 3, analogous to Table \ref{tb:Table2}. With working independence, MB is invalid  and gave inflated type I error rates. All other variance estimators yielded very similar results to those estimating the extended nested exchangeable working correlation structure with MAEE (Table \ref{tb:Table1} and Table \ref{tb:Table2}). In fact, we found out that under balanced designs, the variance estimators BC0, BC1, BC2, AVG, and BC3 are numerically equivalent under either working independence or working extended nested exchangeable correlation matrix (minor differences in empirical type I error and power due to non-convergence of MAEE in a few iterations), when the true correlation structure is extended nested exchangeable. We formally state this result in the following Remark, with the proof presented in Web Appendix I.

\begin{remark}\label{prop2}
Under balanced designs, suppose the randomization is carried out at the $4$th level, and assume arbitrary link and variance functions, with the extended nested exchangeable correlation structure as the true correlation structure. Then, GEE analysis using the extended nested exchangeable working correlation matrix or using an independence working correlation matrix result in the same estimators $\hat\bfbeta$, BC0, BC1, BC2, AVG, and BC3.
\end{remark}

\subsection{Unbalanced designs}
\begin{figure}[htbp]
\centering
\includegraphics[scale=0.49]{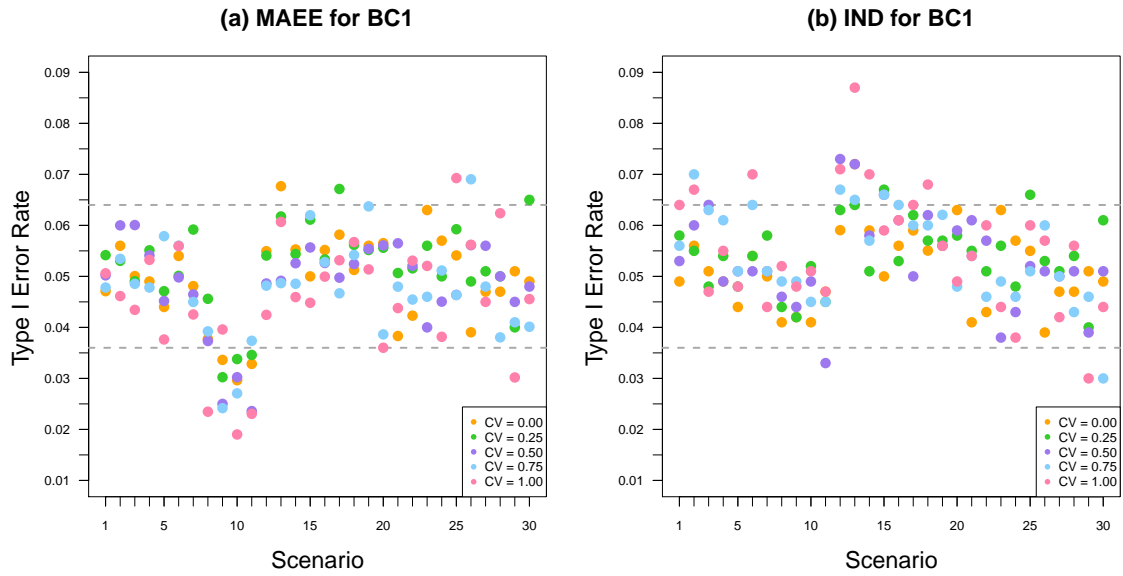}
\caption{Empirical type I error rates of GEE analyses using (a) the extended nested exchangeable working correlation structure with MAEE and (b) an independence working correlation matrix, based on BC1: bias-corrected sandwich estimator of \citet{kauermann2001}, under four-level CRTs.}
\label{fig:Size_BC1}
\end{figure}

\begin{figure}[htbp]
\centering
\includegraphics[scale=0.49]{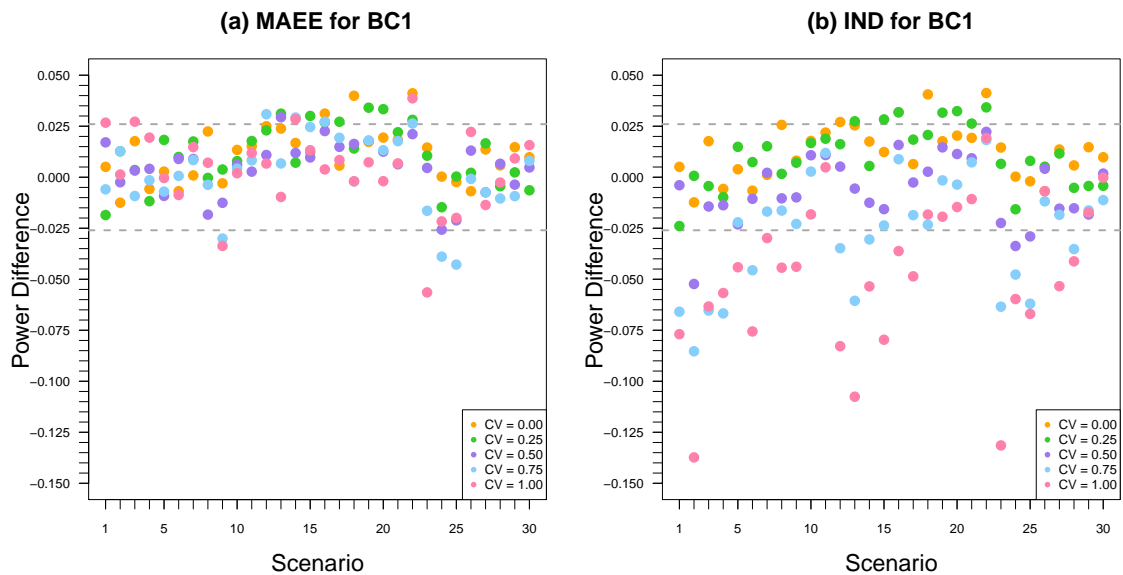}
\caption{Difference between the empirical power and the predicted power of GEE analyses using (a) the extended nested exchangeable working correlation structure with MAEE and (b) an independence working correlation matrix, based on BC1: bias-corrected sandwich estimator of \citet{kauermann2001}, under four-level CRTs.}
\label{fig:Power_BC1}
\end{figure}

Although we derive our sample size formula assuming an equal number of patients within each provider, in practice, providers may have variable numbers of patients. We assessed the robustness of the proposed sample size formula under this specific type of variable panel size, and further illustrate the comparisons between using the true working correlation model versus the independence working correlation model. For each scenario in Section \ref{sec:ecs}, keeping other parameters unchanged, we generated numbers of patients per provider $L_{ijk}$ from a gamma distribution with mean equal to $\bar{L}\in\{5,10\}$ and CV ranging from $\{0.25, 0.50, 0.75, 1\}$. We computed the required sample size ignoring the cluster size variability, but estimate the empirical power under variable cluster sizes for GEE estimators using MAEE with the extended nested exchangeable working correlation structure versus using working independence. The full results for empirical type I error rates and empirical power using different variance estimators are presented in Web Tables 4-11 for GEE analyses using the extended nested exchangeable working correlation structure with MAEE, in Web Tables 12-19 for GEE analyses using working independence, and in Web Figures 1-8 for comparisons of GEE analyses between using the extended nested exchangeable working correlation structure with MAEE and using working independence. As BC1 performed best under balanced designs, here we focus discussion of the results on BC1. \textcolor{black}{Figure \ref{fig:Size_BC1} summarizes the empirical type I error rates for $t$-tests with BC1 across the range of estimated sample sizes, with gray dashed lines indicating the acceptable bounds calculated from a binomial sampling model (detailed calculation of the acceptable bounds is provided in Section \ref{sec:ecs}). Using the extended nested exchangeable working correlation structure with MAEE, most scenarios had valid type I error rates, with liberal results in only five cases across all CV values of cluster sizes (Scenario 13 for CV = 0; Scenarios 17 and 30 for CV = 0.25; Scenario 26 for CV = 0.75; Scenario 25 for CV = 1.00).} On the other hand, as the CV of cluster sizes increases, the GEE estimator with working independence started to exhibit inflated type I error rates. \textcolor{black}{Figure \ref{fig:Power_BC1} summarizes the power results for $t$-tests with BC1, where gray dashed lines indicate the acceptable bounds for difference in empirical versus predicted power (detailed calculation of the acceptable bounds is provided in Section \ref{sec:ecs}). Surprisingly, the empirical power corresponded well with that predicted for most scenarios when the data are analyzed by GEE and MAEE with the extended nested exchangeable working correlation structure, regardless of the cluster size variability; across all CV values of cluster sizes, there were only five cases where the empirical power appeared slightly lower than the predicted (Scenarios 9, 24, and 25 for CV = 0.75; Scenarios 9 and 23 for CV = 1.00).} In sharp contrast, the empirical power for independence GEE had unacceptable lower empirical power than that predicted when the CV of cluster sizes increased from $0.25$ to $1$.

%%%%%%%%%%
% Figure 2
%%%%%%%%%%

%%%%%%%%%%
% Figure 3
%%%%%%%%%%

\section{Applications}\label{sec:app}
\subsection{The RESHAPE trial}
\begin{figure}[htbp]
\centering
\includegraphics[scale=0.38]{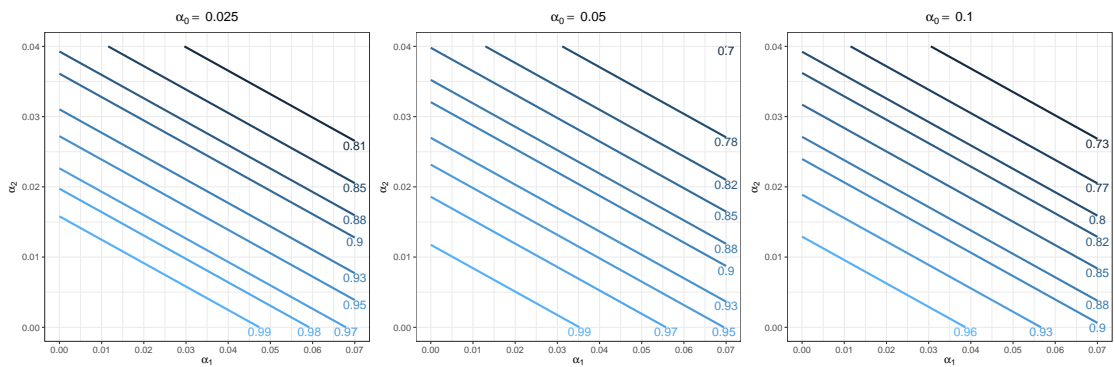}
\caption{Predicted power contours as a function of $\alpha_1$ and $\alpha_2$ at $\alpha_0 = \{0.025, 0.05, 0.1\}$, with $N=22, M=3, K=3, L=36, \text{P}_0=78.5\%, \text{P}_1=88\%$ for the RESHAPE trial, under the logit link function.}
\label{fig:RESHAPE}
\end{figure}

We apply our sample size formula to determine the required sample size in the RESHAPE trial, which is briefly described in Section \ref{sec:intro}. The RESHAPE trial compares two strategies -- RESHAPE and IAU -- for the primary outcome of accuracy of diagnosis of mental illness, defined as the proportion of all patients seen by providers who are accurately diagnosed as determined by psychiatrists. The unit of randomization is the municipality with equal allocation to the two trial arms ($\pi_c=1/2$), and patient-level binary outcomes are collected to measure diagnostic accuracy. The design question focuses on calculating the required number of municipalities to achieve at least 80\% power at the 5\% nominal test size. According to the local conditions in Nepal, we anticipate $3$ health facilities per municipality, $3$ providers per health facility, and roughly $36$ patients per provider. The researchers expect IAU and RESHAPE will result in an accurate diagnosis of $\text{P}_0=78.5\%$ and $\text{P}_1=88\%$, respectively. From extensive discussions with the study team and preliminary simulations for the study protocol, the anticipated ICC between different patients of the same provider is $\alpha_0 = 0.05$, the ICC between different patients of different providers from the same health facility is $\alpha_1 = 0.04$, and the ICC between different patients of different providers from different health facilities in the same municipality is $\alpha_2 = 0.03$. Given $M=3$, $K=3$, $L=36$, and $\bfalpha=(0.05, 0.04, 0.03)$, under the canonical logit link function, Equation \eqref{eq:ssbinOR} and \eqref{eq:power} suggested that the required number of municipalities is $N=22$ with power of $82.65\%$. The same result can also be obtained by using the design effect formula \eqref{eq:de}. Specifically, suppose this is an individually randomized trial without any clustering, then the required number of patients is $562$ (with the working degrees of freedom of a $t$-test adjusted to match the four-level CRT design). By multiplying the design effect $\lambda_4=12.11$, the required number of patients is $6806$ and the required number of municipalities is $6806/(3\times 3\times 36) \approx 22$. Although the research team has good faith in $\alpha_0$, there may be uncertainty in $\alpha_1$ and $\alpha_2$. We conducted a sensitivity analysis by varying the values of these two correlations. Figure \ref{fig:RESHAPE} shows the sensitivity of power as a function of $\alpha_1$ and $\alpha_2$ at $\alpha_0 = \{0.025, 0.05, 0.1\}$, assuming $N=22$ under the logit link function. As expected, the predicted power decreases as $\alpha_0$, $\alpha_1$, or $\alpha_2$ increases. Specifically for $\alpha_0 = 0.05$, power remains above $70\%$ for $\alpha_1\leq 0.07$ and $\alpha_2\leq 0.04$. In Web Appendix L, we provide additional illustrative calculations of the required sample size under the identity and log link functions, as well as when randomization is hypothetically carried out in lower levels under the RESHAPE trial context. Noticeably, if randomization is carried out at the patient level (level 1), the required number of municipalities can dramatically reduce to as few as $6$. 

%%%%%%%%%%
% Figure 4
%%%%%%%%%%

\subsection{The HALI trial}
\begin{figure}[htbp]
\centering
\includegraphics[scale=0.38]{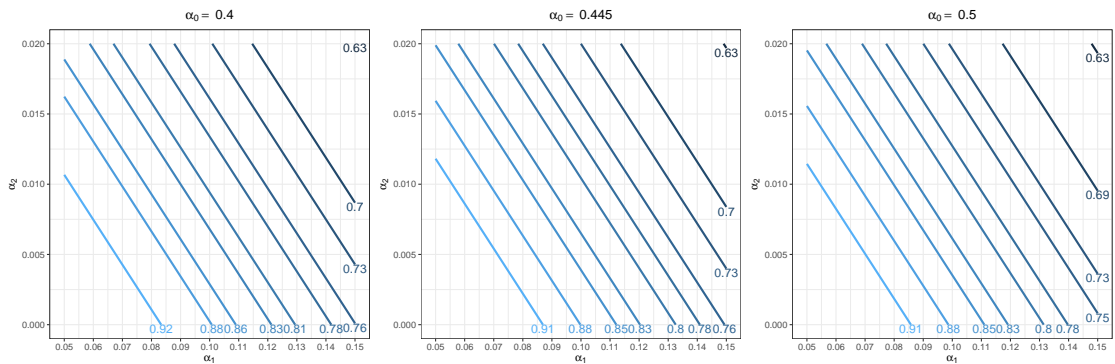}
\caption{Predicted power contours as a function of $\alpha_1$ and $\alpha_2$ at $\alpha_0 = \{0.4, 0.445, 0.5\}$, with $N=36, M=4, K=25, L=2, b=0.19\sqrt{\phi}$ for the HALI trial.}
\label{fig:HALI1}
\end{figure}

We also apply our method to design the HALI trial introduced in Section \ref{sec:intro}. The HALI trial compared two strategies -- the HALI literacy intervention and the usual instruction -- in terms of children's literacy, assessed by various tests such as for spelling and English letter knowledge \citep{jukes2017}. The randomization and implementation were conducted at the TAC tutor zone level with equal randomization ($\pi_c=1/2$), and the first primary outcome of spelling score (ranging from 0 to 20) was measured at the student level. The design focused on calculating the required number of TAC tutor zones to achieve at least 80\% power at the 5\% nominal test size. In the trial sample, there were between 3 and 6 schools in each tutor zone, with 25 children per school, and 2 follow-up visits for literacy tests. For simplicity, we assumed there were 4 schools per TAC tutor zone. According to the results reported in Table 5 of Juke's paper \citep{jukes2017}, the correlation between two spelling scores of each child was $\alpha_0 = 0.445$, the correlation between two spelling scores of different children from the same school was $\alpha_1 = 0.104$, and the correlation between two spelling scores of different children from different schools in the same TAC tutor zone was $\alpha_2 = 0.008$. The researchers initially expected an effect size of 0.19 standard deviation (SD) for spelling scores. Given $M=4, K=25, L=2, b=0.19\sqrt{\phi}$, and $\bfalpha=(0.445, 0.104, 0.008)$, Equation \eqref{eq:sscon} and \eqref{eq:power} suggested that the required number of TAC tutor zones was $N=36$ with power of $80.87\%$. Figure \ref{fig:HALI1} shows the sensitivity of power as a function of $\alpha_1$ and $\alpha_2$ at $\alpha_0 = \{0.4, 0.445, 0.5\}$, assuming $N=36$. Predicted power decreases as $\alpha_0$, $\alpha_1$, or $\alpha_2$ increases. In Web Appendix L, we also provide illustrative calculations of the required sample size under a different effect size, and when randomization is hypothetically carried out in lower levels under the HALI trial context. As expected, the required number of TAC tutor zones reduces to as few as $8$ if randomization is conducted at the children level (level 2). 

%%%%%%%%%%
% Figure 5
%%%%%%%%%%

\section{Discussion}\label{sec:dis}
In this article, we provide a comprehensive investigation on the sample size and power considerations for four-level intervention studies assuming arbitrary link and variance functions, when intervention assignment is carried out at any level, with a particular focus on CRTs. We develop the extended nested exchangeable correlation structure, which is a generalization of the nested exchangeable correlation structure in three-level CRTs \citep{teerenstra2010} and the simple exchangeable correlation structure commonly used in two-level CRTs. \textcolor{black}{That is, the proposed method provides a very general sample size formula that could be applied to continuous, binary and count outcomes in designs with up to four levels.} The results suggest that sample size and power calculations using the proposed method are valid under plausible values of the three correlations ($\alpha_0, \alpha_1, \alpha_2$) for studies with 8 or more clusters. In practice, sensitivity analyses of sample size and power should be performed by varying correlation values within possible ranges, as demonstrated in Section \ref{sec:app}. It is worth noting that the combination of the correlation parameters is valid only if the resulting correlation matrix is positive definite, which can be checked analytically by linear constraints presented in Section \ref{sec:gee}. \textcolor{black}{Finally, although we have primarily focused on four level clustered designs, it is possible to extend our approach to accommodate more than four levels. However, this extension depends on a successful generalization of \eqref{eq:wc} to a more complex correlation matrix and the derivation of its analytical inverse. This interesting extension, however, is beyond the scope of this article and will be pursued in future work.}

In the proposed sample size formula, we assume a balanced design with equal ``cluster" sizes at each level and assume that the analytic model included only an intercept and a \textcolor{black}{binary cluster-level} intervention indicator. At the design stage, assuming the true correlation is extended nested exchangeable, using the model-based variance and using the sandwich variance with an independence working correlation matrix result in the same required sample size for the balanced case. At the analysis stage, using the extended nested exchangeable working correlation structure and using an independence working correlation matrix for GEE analyses result in the same estimators $\hat\bfbeta$, BC0, BC1, BC2, AVG, and BC3 for the balanced case (Theorem \ref{prop1} and Remark \ref{prop2}). However, using the extended nested exchangeable working correlation structure for GEE analyses can protect the study from losing power under unbalanced designs in real-world studies, and allows us to report values of correlations to adhere to the CONSORT Statement \citep{schulz2010, campbell2012}. For such reasons, we recommend modeling the underlying correlation structures and caution the use of independence working correlation matrix for GEE-based design and analysis of multi-level CRTs. 

We compared using the extended nested exchangeable working correlation structure with MAEE and using an independence working correlation matrix for GEE analyses, in our simulation study with $t$-tests. Under balanced designs, BC1 among 7 variance estimators performed best with a near nominal type I error rate and adequate power relative to the proposed sample size formula for as few as 8 clusters using either of these two analysis methods. Similarly, the use of BC1 with a $t$-test was recommended for three-level CRTs with as few as 10 clusters by \citet{teerenstra2010} and for stepped wedge CRTs with as few as 8 clusters by \citet{li2018}. We showed that based on BC1, using the extended nested exchangeable working correlation structure with MAEE, the type I error rate and power maintained acceptable levels when there were variable numbers of level-one units per level-two unit, while using an independence working correlation matrix could not ensure an acceptable type I error rate or power as the CV of numbers of level-one units increased. Therefore, under unbalanced designs, the proposed method of sample size calculations can also be used with the mean number of level-one units provided, and the GEE $t$-test with the use of BC1 can protect the type I error rate and maintain adequate power for four-level CRTs with as few as 8 clusters.

\textcolor{black}{When designing four-level clustered trials in practice, one potential challenge is to identify accurate ICC values for the extended nested exchangeable correlation structure. This challenge, in fact, is not unique to parallel-arm four-level designs, but also equally applies to other CRTs with complex correlation structures (e.g., cluster randomized crossover design and stepped wedge design). Ideally, one could conduct a four-level pilot CRT (e.g., \citet{kim2006multilevel}) to generate ICC estimates. In general, the study planners should search the literature for published ICCs for their outcomes of interest. Depending on the level of clustering in the published trials, only certain components of $\bfalpha$ may be available. In other cases, study planners may also consider using historical or routinely collected data from clusters to estimate $\alpha_0$, $\alpha_1$, $\alpha_2$ and guide the planning of study. When there is a large uncertainty for components of $\bfalpha$, we strongly recommend conducting a sensitivity analysis by considering a range of values for ICCs in power analysis, as we already demonstrated in Section \ref{sec:app}. Finally, we wish to note that publishing ICCs has long been advocated; for example, see \citet{murray2004design} for a table listing 14 research articles presenting ICCs for a variety of endpoints, groups and populations; \citet{preisser2007importance} for published ICCs for nine binary youth alcohol use measures from the Youth Survey of the Enforcing Underage Drinking Laws Program; and \citet{korevaar2021intra} for published ICC estimates from the CLustered OUtcome Dataset bank to inform the design of longitudinal CRTs for a wide range of outcomes. We encourage more such efforts to assist the planning for complex multi-level CRTs and to strengthen the connection between the methodological development on study design and statistical practice.}

\begin{acknowledgement}
This work is funded in part by award R01-MH120649 of the National Institutes of Health for the RESHAPE trial. The funder had no role in study design, data collection and analysis, decision to publish, or preparation of the manuscript. The authors would like to thank the RESHAPE study team for their discussions on the design challenges of this study and, in particular, to Dr. Brandon Kohrt, the Principal Investigator who initiated a research collaboration with us (XW and ELT). We thank the Editor and two anonymous referees for their constructive comments and suggestions, which have improved the exposition of this work.
\end{acknowledgement}
\vspace*{1pc}

\noindent {\bf{Conflict of Interest}}

\noindent {\it{The authors have declared no conflict of interest.}}

\bibliographystyle{biom.bst}
\bibliography{FLCRTbib}

\section*{Supporting Information}
Web Appendices, Tables, and Figures referenced in Sections \ref{sec:intro}-\ref{sec:app}, as well as source code to reproduce the results in Sections \ref{sec:sim}-\ref{sec:app}, are available at \url{https://github.com/XueqiWang/Four-Level_CRT}.

\end{document}